\documentstyle[twocolumn,aps,prl,epsf]{revtex}
\begin{document}
\title{Stripes, pseudogaps, and SO(6) in the cuprate superconductors}

\author{R.S. Markiewicz$^{1,2}$ and M.T. Vaughn$^1$} 

\address{(1) Physics Department and (2) Barnett Institute, 
Northeastern U.,
Boston MA 02115}
\maketitle

\begin{abstract}
We briefly summarize two related calculations.  First, we 
demonstrate that the instabilities (either nesting or pairing) associated with
the high-T$_c$ cuprates can be described by an $SO(6)$ transformation group. 
There are two independent 6-dimensional representations (`superspins'). 
One superspin combines Zhang's 5-component superspin with a flux 
phase instability; the other involves a charge density wave, 
s-wave superconductivity, and an exotic spin current.
\par
The second calculation is a self-consistent slave boson calculation, which
provides a good description of the doping dependence of the photoemission
dispersion in terms of dynamic striped phases.  The stripes are stabilized by
strong electron-phonon coupling, and provide evidence for a doping-dependent
crossover between the two superspin groundstates.
\end{abstract}


\narrowtext

\section{$SO(6)$ Instability Group}

\begin{figure}
\leavevmode
   \epsfxsize=0.44\textwidth\epsfbox{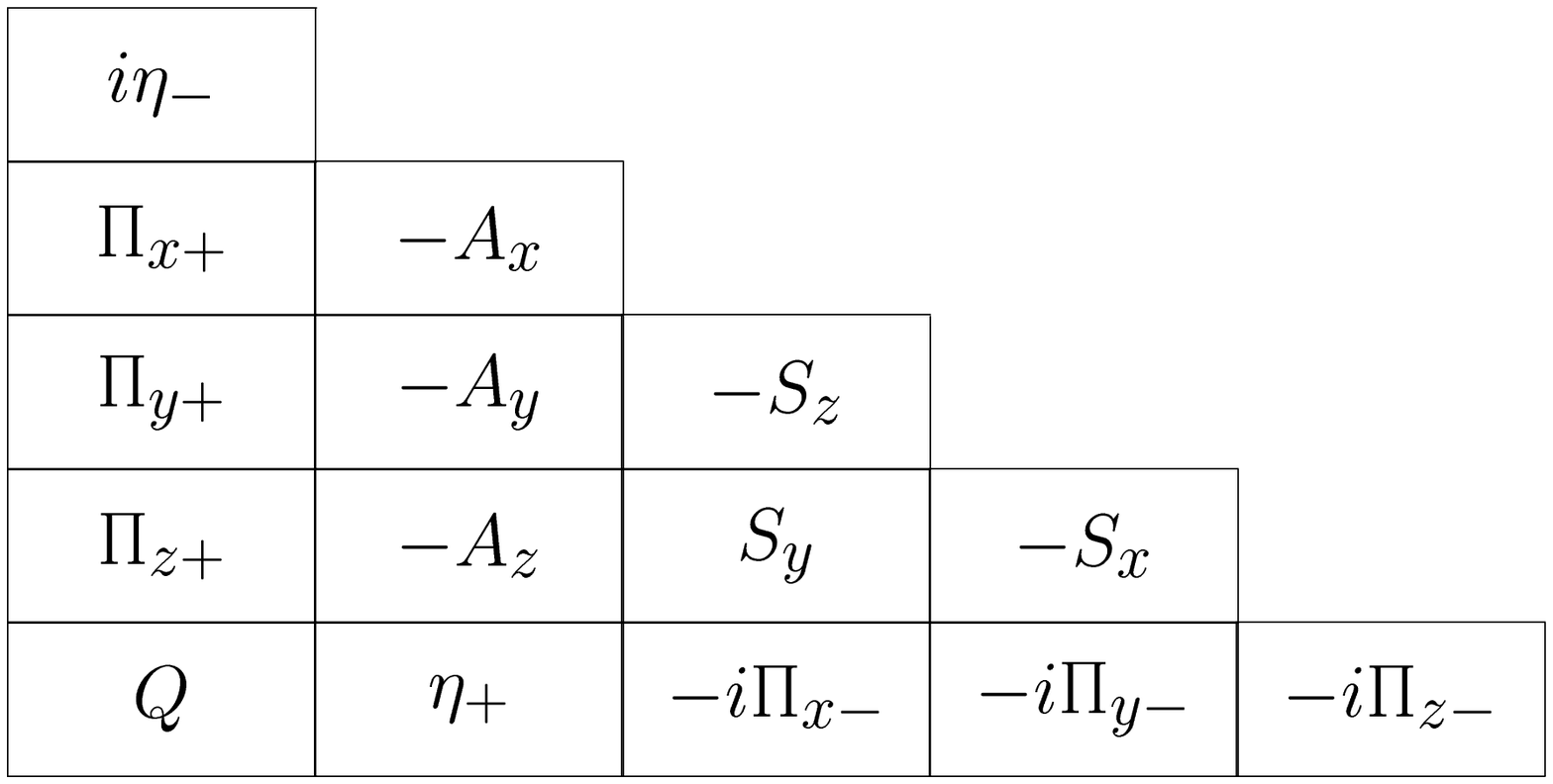}
\vskip-5.50cm 
\caption{Matrix Representation of SO(6), using the shorthand $O_{\pm}=
O\pm O^{\dagger}$.}
\label{fig:0}
\end{figure}
Zhang\cite{Zhang5} has proposed that the physics of doping the cuprate
superconductors is controlled by an $SO(5)$ transformation group, having
a 5-dimensional superspin representation which mixes spin-density wave (SDW)
and d-wave superconducting components.  This is in contradiction to earlier
findings\cite{And} that in the Hubbard model at half filling d-wave
superconductivity is symmetry-equivalent to a flux phase\cite{Affl}.  Here, we
show that Zhang's $SO(5)$ group is a subgroup of an $SO(6)$ group, the
instability group of the two-dimensional (2D) Van Hove singularity 
(VHS)\cite{MarV}.  The corresponding 6-component superspin combines Zhang's
superspin with a flux-phase instability.
\par
Consider an electronic state with a two-fold orbital degeneracy, $\psi_1,\psi_2
$.  Including spin and charge conjugation, 
\begin{figure}
\leavevmode
   \epsfxsize=0.35\textwidth\epsfbox{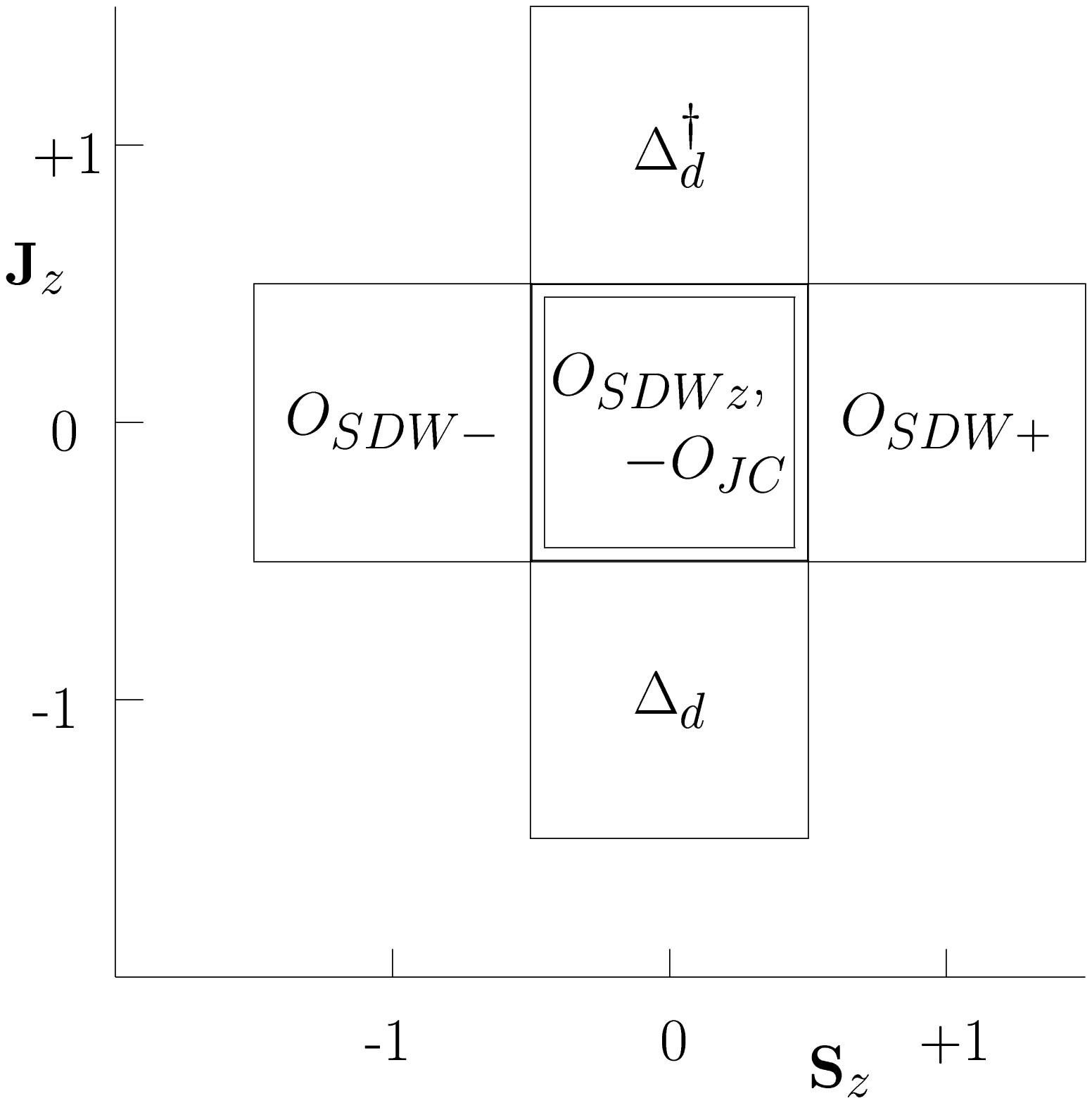}
\vskip-4.5cm 
   \epsfxsize=0.35\textwidth\epsfbox{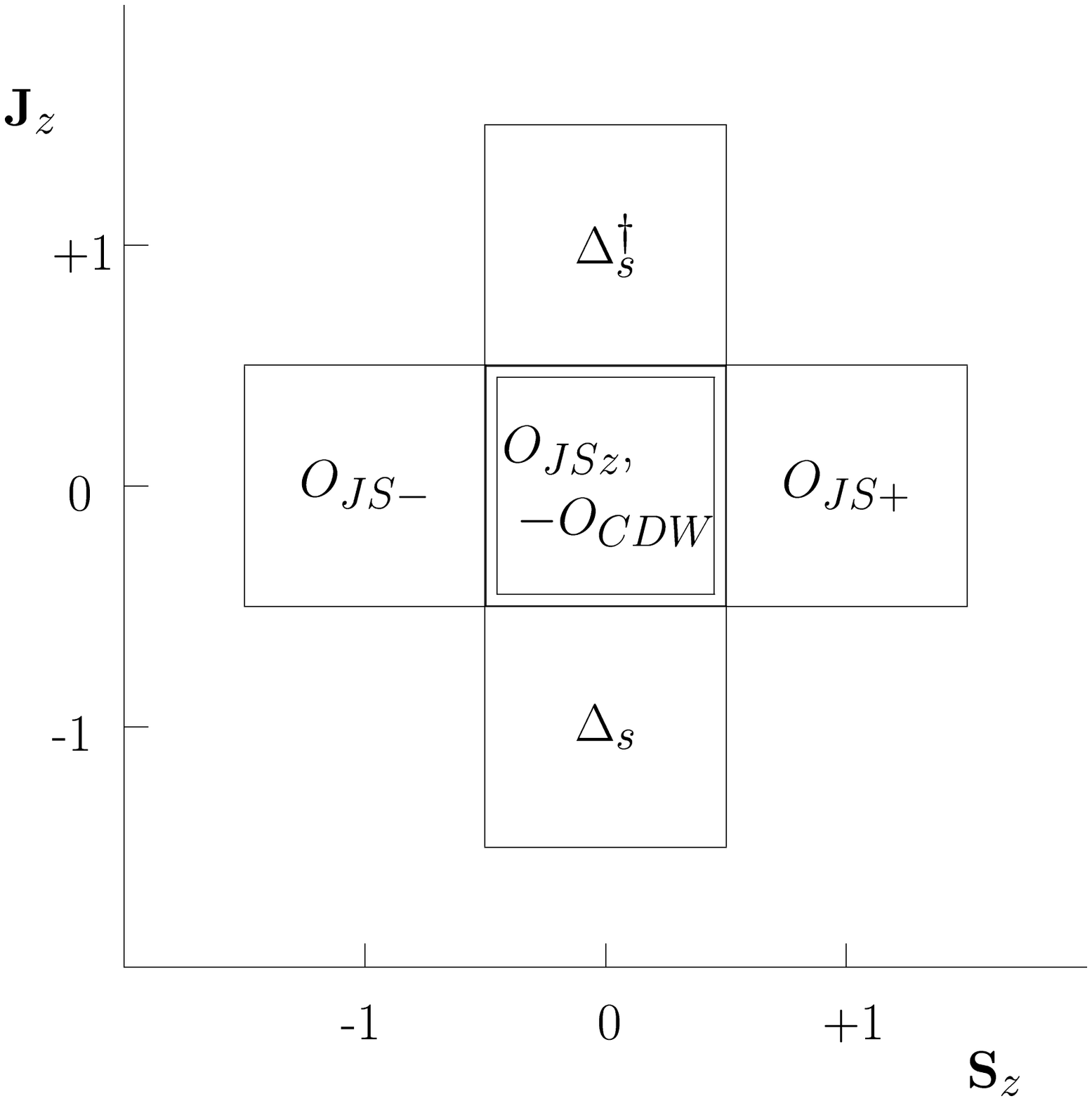}
\vskip-1.6cm 
\caption{$SO(4)$ weight diagram of {\bf V$_s$} (top) and {\bf V$_h$} (bottom), 
where $S_z$ ($J_z$) is the z component of spin (pseudospin).}
\label{fig:1}
\end{figure}
\par\noindent
8 single particle operators can be
formed, or $8\times 7/2=28$ pair operators (taking account of the Pauli
exclusion principle).  These operators can be combined into a representation of
an $SO(8)$ group, the pair group\cite{MarV} which controls the various pairing 
and nesting instabilities which can arise.  For the quasi-2D cuprates, the
orbital degeneracy is associated with the VHS's along $(\pi ,0)$ and $(0,\pi )$;
because of the large density of states (dos) at a VHS, the Fermi surface can be
approximated by just two points\cite{Dzy,Sch2}, similar to a one-dimensional 
metal.  
\par
Consider a supercell of size $2a\times 2a$, each cell containing four 
atoms.  Let $\vec i$ be a lattice vector of the original lattice, and $\vec r$ 
be a lattice vector of the supercell lattice.  If the Cu-site operators are
$a$'s, then the VHS operators are
\begin{eqnarray}
\psi_{1\sigma}(\vec r)={1\over 2}(a_{\vec i,\sigma}-a_{\vec i+\hat x,\sigma}
+a_{\vec i+\hat y,\sigma}-a_{\vec i+\hat x+\hat y,\sigma})\nonumber \\
\psi_{2\sigma}(\vec r)={1\over 2}(a_{\vec i,\sigma}+a_{\vec i+\hat x,\sigma}
-a_{\vec i+\hat y,\sigma}-a_{\vec i+\hat x+\hat y,\sigma}).
\label{eq:A1}
\end{eqnarray}
(The remaining operators are located in other parts of the Brillouin
zone, and play a negligible role in nesting phenomena.)
\par
A special role is played by the operator $\tau =\sum_{\vec r,\sigma}(
\psi^{\dagger}_{1\sigma}(\vec r)\psi_{1\sigma}(\vec r)-\psi^{\dagger}_
{2\sigma}(\vec r)\psi_{2\sigma}(\vec r))$, which splits the VHS degeneracy.
The operators which commute with $\tau$ form an $SO(6)$ transformation group,
Fig. 1, with two fundamental 6-dimensional representations, Fig. 2.  The
momentum-space representation of the superspin operators is shown in Table I;
the full list is in Ref. \cite{MarV}.  Since the generators of $SO(6)$ can be 
represented as antisymmetric $6\times 6$ matrices, Fig. 1 assigns each operator 
to the position of the corresponding non-zero matrix element, to produce the 
correct $SO(6)$ Lie algebra
\begin{equation}
[L^{ij},L^{km}]=i(\delta_{ik}L^{jm}+\delta_{jm}L^{ik}-\delta_{im}L^{jk}-
\delta_{jk}L^{im}).
\label{eq:1}
\end{equation}
Note that by deleting the second row and column, an $SO(5)$ subalgebra is
generated, which is equivalent to the one postulated by Zhang\cite{Zhang5}.
\par
In Fig. 2, the pseudospin operator\cite{Zhang4} is 
\begin{equation}
J_-=2\eta,\ \ J_+=2\eta^{\dagger},\ \ J_z=Q,
\label{eq:8}
\end{equation}
and $O_{SDW\pm}=\mp(O_{SDWx}\pm iO_{SDWy})/\sqrt{2}$.
The superspin ${\bf V_s}$ consists of a spin-density wave spin triplet ($\vec
O_{SDW}$) and a pseudospin triplet combining d-wave superconductivity ($\Delta_
d$,  $\Delta_d^{\dagger}$) and an orbital antiferromagnet ($O_{JC}$) equivalent 
to the flux phase operator\cite{Affl}.  The superspin ${\bf V_h}$ consists of a
spin current spin triplet ($\vec O_{JS}$) and a charge-density wave ($O_{CDW}$),
s-wave superconductivity ($\Delta_s$, $\Delta_s^{\dagger}$) pseudospin triplet.
These operators are equivalent to those
introduced by Schulz\cite{Sch2} and, in a related context, by Halperin and
Rice\cite{HalR}.  
\vskip 0.2in
\begin{tabular}{||c|c||}        
\multicolumn{2}{c}{{\bf Table I: Fourier Transforms of Superspins}} \\ 
            \hline\hline
Operator & Representation \\   
    \hline\hline
$O_{CDW}$ & $2\Sigma_{\vec k\sigma}(a^{\dagger}_{\vec k\sigma}a_{\vec k+\vec Q,
 \sigma})$   \\     \hline
$O_{SDW\alpha}$ & $2\Sigma_{\vec k,i,j}(a^{\dagger}_{\vec k,i}
 \sigma_{ij}^{\alpha}a_{\vec k+\vec Q,j})$   \\     \hline
$\Delta_s$ & $2\Sigma_{\vec k}(a_{\vec k\uparrow}a_{-\vec k\downarrow})$
  \\     \hline
$O_{JC}$ & $-\Sigma_{\vec k\sigma}g(\vec k)(a^{\dagger}_{\vec k\sigma}a_{\vec k+
 \vec Q,\sigma})$   \\     \hline
$O_{JS\alpha}$ & $-\Sigma_{\vec k,i,j}g(\vec k)(a^{\dagger}_{\vec k,i}
 \sigma_{ij}^{\alpha}a_{\vec k+\vec Q,j})$   \\     \hline
$\Delta_d$ & $-\Sigma_{\vec k}g(\vec k)(a_{\vec k\uparrow}a_{-\vec k\downarrow})$
  \\     \hline
\end{tabular}
\vskip 0.2in
\par
In weak coupling, the instabilities are controlled by their respective
susceptibilities and the form of the quartic interaction terms.  For an energy 
dispersion
\begin{equation}
\epsilon_{\vec k}=-2t(\cos k_xa+\cos k_ya)+4t^{\prime}\cos k_xa\cos k_ya,
\label{eq:24}
\end{equation}
then when $t^{\prime}=0$ and at exactly half filling (square Fermi surface), the
model has an extra pseudospin symmetry\cite{Zhang4}, and all the bare
susceptibilities are equal\cite{MarV}.  The dressed susceptibilities are
\begin{equation}
\chi_i={\chi_{0i}\over 1+2\chi_{0i}G_i},
\label{eq:22}
\end{equation}
where $G_i$ is related to the quartic coupling.  Since all 12 $\chi_{0i}$'s are
equal (at half filling), the ground state instability is controlled by the 
interactions $G_i$.  For a pure Hubbard model, the $U$ term breaks the $SO(6)$ 
symmetry, leading to a negative $G_{SDW}$ and a positive $G_{CDW}$, $G_s$,
Fig. 3.  Thus at half filling the ground state is antiferromagnetic, $O_{SDW}$.
\begin{figure}
\leavevmode
   \epsfxsize=0.50\textwidth\epsfbox{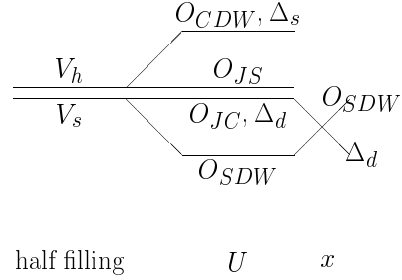}
\vskip-6.5cm 
\caption{Splitting of $SO(6)$ superspin degeneracies by interaction $U$ and by
doping $x$.}
\label{fig:2a}
\end{figure}
\par
Doping away from half filling breaks the pseudospin symmetry, leading to two
distinct susceptibilities, $\chi_{00}$ for the particle-hole operators 
and $\chi_{02}$ for the particle-particle operators, Fig. 4.  Since the
susceptibilities are enhanced by nesting, their magnitudes fall off with doping.
However, the magnitude of $\chi_{00}$ falls much more rapidly, leading to a
possible scenario for d-wave superconductivity, Fig. 3\cite{MarV}: if there is
an attractive interaction for d-wave superconductivity, it will be overwhelmed
by the strong divergence in $\chi_{SDW}$ at half filling, but with doping, the
SDW instability falls off much faster, leading to a possible crossover to
d-wave superconductivity.  This is consistent with earlier renormalization 
group results\cite{Dzy,Sch,Surv}. 
\par
In a more realistic, strong coupling model, a number of additional features
must be considered.  First, additional hopping parameters, such as the 
second-neighbor $t^{\prime}$ term in Eq.~\ref{eq:24}, also break the
pseudospin degeneracy, Fig. 5.  Note that the susceptibility diverges away from 
half filling, at the VHS, but even at optimal doping $\chi_{02}>>\chi_{00}$.
Moreover, by comparing Fig. 4 it can be seen that $\chi_{02}$ is actually
larger, at optimal doping, for $t^{\prime}\ne 0$.  Since strong coupling pins
the Fermi level near the VHS over an extended doping range\cite{RM3,Pstr}, this
enhancement of $\chi_{02}$ can greatly enhance the possibility for d-wave
superconductivity to arise on doping.
\par
There is considerable evidence for phonon coupling and structural 
instabilities in the doped material\cite{Surv}.  Hence, it is interesting to 
see whether strong electron-phonon coupling can cause a crossover to a 
groundstate involving {\bf V$_h$}.  Since the CDW operator is proportional to
the difference in hole population of the odd and even sublattices, it must 
vanish in the strong coupling limit at 
\begin{figure}
\leavevmode
   \epsfxsize=0.33\textwidth\epsfbox{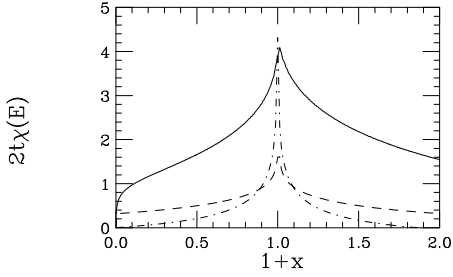}
\vskip0.5cm 
\caption{Susceptibilities $\chi_{00}$ (dotdashed line) and $\chi_{02}$ (solid
line) and density of states (dashed line) vs. band filling $1+x$ for the one 
band model with $t^{\prime}=0$.}
\label{fig:2}
\end{figure}
\begin{figure}
\leavevmode
   \epsfxsize=0.33\textwidth\epsfbox{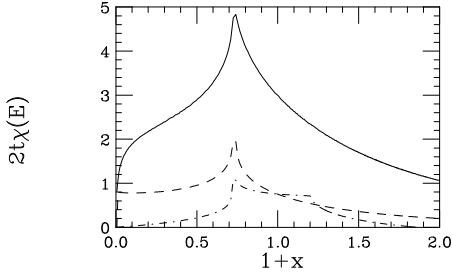}
\vskip0.5cm 
\caption{As in Fig.~\protect\ref{fig:2}, but with $2t^{\prime} =-0.3t$.}
\label{fig:3}
\end{figure}
\par\noindent
half filling, when there is one 
hole per Cu.  For finite doping there can be a CDW order, with $<O_{CDW}>
\propto x$, if all the holes are confined to one sublattice.  The resulting 
structure is equivalent to that found in La$_{2-x}$Sr$_x$NiO$_4$ for $x\sim 
0.5$\cite{ZL}.  Moreover, such a structure is closely related to the holon
condensate\cite{any} introduced to overcome parity-violation problems in 
anyonic models of superconductivity.  In these models, two species of holons are
introduced, of opposite parity violation. A coherent superposition of the two 
preserves parity, while producing a {\it holon condensation}, wherein all holons
are confined to one sublattice, leading to a real-space CDW\cite{SWij}.
\par
In the following section, we describe a slave boson calculation\cite{Pstr} which
self-consistently takes into account both strong coupling effects and strong
electron-phonon coupling.  We show that VHS pinning can provide a natural
explanation for the observed stripe phases\cite{Tran,Dai} in these materials, as
well as pseudogaps and extended VHS's.

\section{CDW Order and Phase Separation}

In the strong coupling limit, the dominant physics is quite different.  At
exactly half filling, there is a metal-insulator transition that is unrelated to
nesting: a Mott transition, or charge transfer insulator transition in the
three-band model of the cuprates which will be employed here.  The 
nesting/pairing instabilities arise on top of this transition, and are strongly
constrained by the large Hubbard $U$, which limits double occupancy of the Cu
orbitals.  In the slave boson approach employed here, the limit $U\rightarrow
\infty$ is taken (no double occupancy).  To include spin effects, it is 
necessary to go beyond the standard slave boson approach, either by introducing
four slave bosons\cite{KoR} or by introducing a Heisenberg exchange 
$J$\cite{DiDo} -- the latter approach is followed here.

Thus, the starting point of the analysis is the three-band model Hamiltonian: 
\begin{eqnarray}
H=\sum_{j,\sigma}\bigl(\Delta d^{\dagger}_{j,\sigma} d_{j,\sigma}
+\sum_{\hat\delta}
t_{CuO}[d^{\dagger}_{j,\sigma}p_{j+\hat\delta,\sigma}
+(c.c.)] \nonumber \\
+\sum_{\hat\delta^{\prime}}t_{OO}[p^{\dagger}_{j+\hat\delta,\sigma}p_{j+
\hat\delta^{\prime},\sigma}+(c.c.)] 
\nonumber \\
+J\sum_{\hat\delta^{\prime\prime},\sigma^{\prime}}
d^{\dagger}_{j,\sigma}d_{j,\sigma^{\prime}}
d^{\dagger}_{j+\hat\delta ,\sigma^{\prime}}d_{j+\hat\delta ,\sigma}
\bigr).
\label{eq:1a}
\end{eqnarray}
In this equation,  
$d^{\dagger}$ ($p^{\dagger}$) is a creation operator for holes on Cu (O),
j is summed over lattice sites, $\hat\delta$ over Cu-O nearest neighbors, 
$\hat\delta^{\prime}$ ($\hat\delta^{\prime\prime}$) over O-O (Cu-Cu) neighbors, 
and c.c. stands for complex conjugate.  
The band parameters are $\Delta$, the splitting between the Cu and O 
energy levels, $t_{CuO}$, the Cu-O hopping parameter, $t_{OO}$, the O-O 
hopping parameter, and $J$ the exchange constant.
\par
In the above equation, the limit $U\rightarrow\infty$ has been taken, and a
slave boson introduced to prevent Cu double occupancy.  The parameters $t_{CuO}$
and $\Delta$ are renormalized values, to be solved self-consistently\cite{Pstr}.
In the exchange energy, the $d^4$-term is decoupled in mean-field as
\begin{eqnarray}
J\sum_{j,\hat\delta^{\prime\prime},\sigma ,\sigma^{\prime}}
d^{\dagger}_{j,\sigma}d_{j,\sigma^{\prime}}
d^{\dagger}_{j+\hat\delta ,\sigma^{\prime}}d_{j+\hat\delta ,\sigma}
\rightarrow \nonumber \\
{2N_s\over J}\Delta_1^2+\sum_{j,\hat\delta^{\prime\prime},\sigma}
(\Delta_{j,j+\hat\delta^{\prime\prime}}d^{\dagger}_{j,\sigma}d_{j+\hat\delta^
{\prime\prime},\sigma}+ c.c.)
\label{eq:2}
\end{eqnarray}
where $N_s$ is the number of unit cells and
\begin{equation}
\Delta_{ij}\equiv J\sum_{\sigma}<d_{i\sigma}d_{j\sigma}^{\dagger}>
=\Delta_1e^{i\theta_{ij}}.
\label{eq:13d}
\end{equation}
Depending on the choice of phase, a variety of magnetic states are possible.  
Here the only choices considered are $\theta_{ij}=0$, corresponding to a 
uniform phase (only short-range magnetic correlations), and $\theta_{ij}=\pm
\pi /4$ for a flux\cite{Affl} phase, with the $\pm$ sign chosen so that
the net phase change around any plaquette is $\pi$.  
\par
Finally, electron-phonon coupling is included by the substitution $t_{CuO}=t
(1\pm\delta )$, with $\delta$ due to a CDW distortion: for a breathing mode, all
four Cu-O bonds are equivalent for any given Cu, but all are long ($t_{CuO}=t
(1-\delta )$) for one sublattice, and short for the other.  Note that this model
includes only two of the twelve possible superspin components, but already gives
rise to a complicated phase diagram.  It should describe the ${\bf V_s
\rightarrow V_h}$ ($O_{JC}\rightarrow O_{CDW}$) crossover.
\par
\begin{figure}
\leavevmode
   \epsfxsize=0.33\textwidth\epsfbox{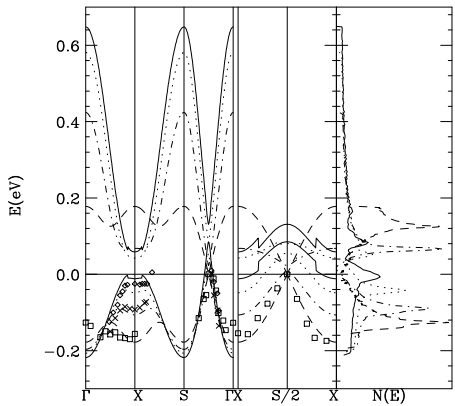}
\vskip0.5cm 
\caption{(a) Energy dispersion for the flux phase at half filling $x_1=0^+$ 
(dashed line) and the paramagnetic phase at $x=$ 0.22, V$_{ep}$ = 0.6eV (solid 
line), close to the minimum free energy, and for the fluctuating stripe phase 
model, for $\nu_c$ = 0.5 (dotdashed line) and 0.75 (dotted line).
Data from underdoped Bi-2212 (diamonds and 
$\times$'s)\protect\cite{Gp0} or SCOC (squares)\protect\cite{Well} are plotted 
as E/2.  Special points of the Brillouin zone are X = ($\pi$,0), S = ($\pi ,\pi
$). (b) Density of states vs. energy, for the same parameters.}
\label{fig:5}
\end{figure}

A charge transfer insulator transition is found at half filling when $\Delta$
exceeds a critical value\cite{Cast}.  Near half filling, the flux phase is 
stable, and the CDW is suppressed by strong coupling effects.  However, if the 
electron-phonon coupling is strong enough, there is a crossover in the doped
material to a CDW phase, which is maximally stable near the doping at which the
Fermi level crosses the bare VHS.  As had been previously noted\cite{Laugh},
the flux phase provides a good description of the measured energy dispersion
in insulating cuprates\cite{Well}, Fig.~\ref{fig:5}, while the CDW fits the
dispersion near optimum doping\cite{Gp0}, in particular reproducing the
extended VHS.  [Note however that the present calculations underestimate the
total dispersion by a factor of $~$2.]
\par
The situation for intermediate dopings is more complex.  The free energy minimum
at finite doping associated with the CDW leads to an instability toward phase
separation between magnetic (flux phase) and charged (CDW) domains.  
There is considerable experimental evidence for such phase separation in the 
cuprates\cite{phassep,Eme1,Surv}, which can be macroscopic if the doping
counterions are mobile (e.g., the interstitial oxygen in La$_2$CuO$_{4+\delta}
$\cite{LCO}.  More commonly, Coulomb repulsion between holes restricts the phase
separation to a nanoscopic scale\cite{Nag}, typically forming striped 
phases\cite{Tran,Dai}.  The driving mechanism for this phase separation has
variously been ascribed to large $J$\cite{Eme,HM}, to the nearest neighbor 
Coulomb energy $V$\cite{Rai}, or, as in the present case, to VHS-induced CDW
formation\cite{RM3}.  
\par
\begin{figure}
\leavevmode
   \epsfxsize=0.33\textwidth\epsfbox{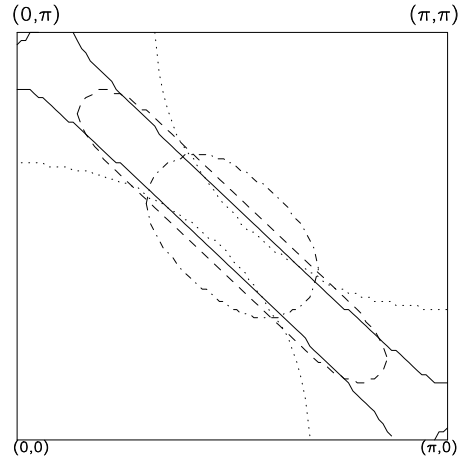}
\vskip0.5cm 
\caption{Fermi surfaces for the fluctuating stripe phase model, for $\nu_c$ = 
1.0 (dotted line), 0.5486 (solid line), 0.5 (dashed line), and 0.25 (dotdashed 
line).}
\label{fig:11}
\end{figure}

In modeling the striped phase, a static stripe pattern produces too much 
structure in the dispersion, due to minigaps\cite{SEK,Pstr}.  If a fluctuating
striped phase is approximated by taking the weighted average of the end-phase
band parameters, good agreement with the experimental dispersion is found, 
Fig.~\ref{fig:5}.  In this figure $\nu_c=x/x_c$ is the fraction of charged 
stripes, with $x_c$ the doping of the uniformly doped charged end phase, here
taken as $x_c=0.288$.  The same model also describes the crossover from small
(hole pockets) to large Fermi surface with doping, Fig.~\ref{fig:11}.
\par
A number of points should be made.  (1) The splitting of the bands near the
$X$-point, $(\pi ,0)$, provides a good descrpition of pseudogap formation in the
cuprates -- not only the photoemission data, Fig.~\ref{fig:5}, but also 
thermodynamic data from heat capacity and magnetization\cite{Gp3}, as well as
the temperature-doping phase diagram\cite{RMPRL}.  The smooth doping dependence
masks a crossover from spin-like to charge-like instability -- from ${\bf V_s
\rightarrow V_h}$.  Accordingly, the underdoped results near half filling are in
good agreement with recent Monte Carlo calculations\cite{PHGE}.  (2) From 
Fig.~\ref{fig:5}b, it is clear that the pseudogap is directly associated with
the splitting of the VHS degeneracy, and hence with the properties of $SO(6)$.
(3) Campuzano\cite{Camp} reports a different doping dependence of the $(\pi ,0)$
photoemission, namely that there are two independent features, one a sharp 
quasiparticle peak which is near the 
Fermi level at optimal doping, and broadens severely in underdoped samples, and 
the second a broad peak which is already present at $\sim$200meV below the Fermi
level in optimally doped material, and gradually shifts to 300meV with increased
underdoping.  A dynamic fluctuation model can accomodate this in terms of a
dynamic average of the two end phases. 
(4) There is considerable evidence for short-range CDW order in the doped 
cuprates, summarized in Ref. \cite{Surv}, Section 9.2, often associated with 
CuO bond stretching\cite{RMXB}.  
(5) In the present model, the pseudogap is associated with magnetic or
structural instabilities which compete with superconductivity. Hence, there
should be a separate superconducting gap, which scales with $T_c$ and 
$\nu_c$\cite{RMPRL}.  This gap may be related to the spin gap and resonance peak
seen in neutron scattering measurements of the magnetic susceptibility near 
$S=(\pi ,\pi )$\cite{spgp}.  
(6) In the overdoped regime, the CDW undergoes a quantum phase transition, 
$T_{CDW}\rightarrow 0$.  It has been suggested that such a QCP can lead to
high-T$_c$ superconductivity\cite{PCCG}.  
\par
MTV's research is supported by the Dept. of Energy under Grant \#
DE-FG02-85ER40233.  Publication 711 of the Barnett Institute.

\end{document}